\documentclass[12pt]{article}
\usepackage{amsmath,amssymb,epsfig}
\pdfoutput=1
\usepackage{cite}

\usepackage{color}
\input{colordvi.tex}
\def\unit{{\relax{\rm 1\kern-.26em I}}}

\makeatletter

\renewcommand\section{\@startsection {section}{1}{\z@}%
                                   {-3.5ex \@plus -1ex \@minus -.2ex}%
                                   {2.3ex \@plus.2ex}%
                                   {\normalfont\large\bfseries}}

\renewcommand\subsection{\@startsection{subsection}{2}{\z@}%
                                     {-3.25ex\@plus -1ex \@minus -.2ex}%
                                     {1.5ex \@plus .2ex}%
                                     {\normalfont\normalsize\bfseries}}

\makeatother

\begin{document}

\baselineskip=18pt  
\numberwithin{equation}{section}  
\allowdisplaybreaks  



%
%


\thispagestyle{empty}

\vspace*{-2cm}
\begin{flushright}
\end{flushright}

\begin{flushright}

\end{flushright}

\begin{center}

\vspace{1.4cm}

\vspace{1cm}
{\bf\Large Low-Scale Gauge Mediation with a 100 TeV Gravitino
}
\vspace*{1.3cm}

{\bf
Masaki Asano$^{1}$, Yuichiro Nakai$^{2}$ and Norimi Yokozaki$^{3}$} \\
\vspace*{0.5cm}

${ }^{1}${\it Physikalisches Institut and Bethe Center for Theoretical Physics, Universit\"at Bonn, Nussallee 12, D-53115 Bonn, Germany}\\
${ }^{2}${\it Department of Physics, Harvard University, Cambridge, MA 02138, USA}\\
${ }^{3}${\it Istituto Nazionale di Fisica Nucleare, Sezione di Roma, Piazzale Aldo Moro 2, I-00185 Rome, Italy}\\

\vspace*{0.5cm}

\end{center}

\vspace{1cm} \centerline{\bf Abstract} \vspace*{0.5cm}

We propose a new framework of low-scale gauge-mediated supersymmetry (SUSY) breaking
with a gravitino mass of $\mathcal{O}(100) \, \rm TeV$.
The usual 4D gauge mediation models predict a light gravitino and suffer from cosmological problems. 
In our framework, a heavy gravitino in gauge mediation is realized with a flat extra-dimension, whose compactification scale is around the grand unified theory scale. Superparticle masses of the visible sector from gravity/anomaly mediation are suppressed, and they are purely generated by usual gauge mediation on the visible brane.
Importantly, the Higgs $B_\mu$-term vanishes at the leading order, 
which enables us to obtain the suitable $\mu$-$B_\mu$ relation for the electroweak symmetry breaking.
We discuss such models considering two possibilities of the SUSY breaking source: 1) Scherk-Schwarz SUSY breaking which we call  Scherk-Schwarz Gauge Mediation and 2) gravitational SUSY breaking localized on a hidden brane. 
In the case 2), the cosmological moduli problem may be relaxed as well.

\newpage
\setcounter{page}{1} 



\section{Introduction}\label{sec:intro}

Gauge mediation is an attractive mechanism to mediate supersymmetry (SUSY) breaking into ordinary superparticles \cite{gmsb, gmsb_old}
(for reviews, see e.g. \cite{Giudice:1998bp,Kitano:2010fa}). 
The mediation is achieved through the standard model (SM) gauge interactions. Since the SM gauge interactions are flavor blind, it can suppress dangerous flavor changing neutral currents (FCNCs).
Gravity mediation effects do not spoil this advantage if the mediation scale is much smaller than the Planck scale. The gravitino mass is, therefore, typically smaller than the weak scale in the gauge mediation scenario. However, the gravitino with mass $m_{3/2} > \mathcal{O}(10) \, \rm eV$ can raise cosmological problems such as over-closure of the universe \cite{Pagels:1981ke}.
Only low-scale SUSY breaking with a gravitino mass as light as $1$-$16 \, \rm eV$
is allowed~\cite{Viel:2005qj}, but in this case, we encounter the constraint from vacuum instability in general
\cite{Hisano:2007gb,Hisano:2008sy},\footnote{ For general arguments on the relation between the gaugino mass and the vacuum structure, see \cite{Komargodski:2009jf,Nakai:2010th}. } and most cases are already excluded by SUSY searches at the Large Hadron Collider (LHC).

In this paper, we propose a new framework of gauge mediation models in which the gravitino mass is $\mathcal{O}(100) \, \rm TeV$.\footnote{
Gauge mediation with a gravitino mass at the weak scale was presented in \cite{Nomura:2000uw,Nomura:2001ub}.
Reference~\cite{Goh:2005be} discussed a possibility of gauge mediation with a heavy gravitino in the context of the emergent SUSY
\cite{Sundrum:2009gv,Gherghetta:2011wc,Heidenreich:2014jpa,Nakai:2014iea}
where the standard cosmology is significantly affected by a warped extra dimension. See also \cite{Murayama:2007ge,Shirai:2008qt,Craig:2008vs,Ibe:2009pq,Ding:2013pya}.
}
Since such a heavy gravitino decays before the Big Bang Nucleosynthesis (BBN), it does not spoil the success of the BBN and can address the gravitino problem.
In order to achieve such a mass spectrum, we introduce a flat extra dimension, a $S^1 / Z_2$ orbifold, as small as the unification scale. All the SM multiplets as well as the messenger sector fields are localized on a brane at the fixed point. On the other hand, a hidden gauge multiplet propagates in the entire 5 dimensional (5D) space. As the SUSY breaking source, we discuss two possibilities: 1) Scherk-Schwarz SUSY breaking and 2) gravitational SUSY breaking localized on a hidden brane.

In the first scenario which we call {\it Scherk-Schwarz Gauge Mediation} (SSGM), 
SUSY breaking is provided by the Scherk-Schwarz mechanism
\cite{Scherk:1978ta}.\footnote{
Scherk-Schwarz SUSY breaking models
have been studied in e.g. \cite{Pomarol:1998sd,Antoniadis:1998sd,Barbieri:2000vh,Barbieri:2001yz,Barbieri:2001dm}.
For more recent works, see \cite{Murayama:2012jh,Kitano:2012cz,Dimopoulos:2014aua,Tobioka:2015vsv}.
}
Since a hidden gauge multiplet propagates in the entire 5D space, the hidden gaugino mass is generated at the gravitino mass scale.
%
The SUSY breaking in the messenger sector is generated by radiative corrections via hidden gauge charged fields on the visible brane.

The point is that the $F$-term of the so-called chiral compensator field is much smaller than the gravitino mass \cite{Marti:2001iw,Luty:2002hj}. Thus, anomaly mediation effects \cite{Randall:1998uk,Giudice:1998xp} are tiny in this scenario.
The Higgs $B_\mu$-term vanishes at the leading order and the correct $\mu$-$B_\mu$ relation
for the electroweak symmetry breaking (EWSB) can be obtained~\cite{cp-safe}. As a result, the soft masses of ordinary superparticles are simply generated by gauge mediation while the gravitino is very heavy.
The mass spectrum of the minimal supersymmetric standard model (MSSM) is the same as the usual gauge mediation spectrum except the gravitino mass. In general cases of the messenger sector or the Higgs sector, the lightest supersymmetric particle (LSP) can be the weakly interacting massive particle (WIMP) and a candidate of the dark matter.

Although we consider the Scherk-Schwarz mechanism in the first scenario, it is not necessarily required to achieve such sparticle mass spectra. Actually, in the second scenario, we consider gravitational SUSY breaking~\cite{kiy} on the hidden brane and show the similar spectra
as the first scenario. A shift symmetry of the SUSY breaking field~\cite{cp-safe} forbids dependence on the SUSY breaking field  in the superpotential, and the compensator $F$-term is still zero for the vanishing cosmological constant, as in the Scherk-Schwarz case.

The rest is organized as follows. In the next section, we present a model of the SSGM and estimate the mediated SUSY breaking in the visible sector. The brane-localized SUSY breaking case is considered in section $3$. Section $4$ is devoted to conclusions and discussions.

\section{Scherk-Schwarz Gauge Mediation}\label{sec:model}

We consider a flat 5D space whose fifth dimension is compactified on an $S^1/ Z_2$ orbifold: $0 \leq y \leq 2\pi R$
with identification $y \leftrightarrow -y$.
The 5D Planck scale $M_5$ is related to the 4D Planck scale $M_4$ as $M_4^2 = 2 \pi R M_5^3 = L M_5^3$, leading to
\begin{eqnarray}
M_5 \approx 3.9 \times 10^{17} {\rm GeV} \left(\frac{L^{-1}}{10^{16}\,{\rm GeV}}\right)^{1/3}.
\end{eqnarray}
There are branes at each of the fixed points, $y = 0, \pi R$.
The bulk of the extra dimension respects 5D $\mathcal{N}=1$ SUSY which is $\mathcal{N} = 2$ in 4D sense.
At the fixed points, only 4D $\mathcal{N} = 1$ SUSY is realized.
We introduce a hidden $SU(N)$ gauge field propagating in the entire 5D space,
a 4D $\mathcal{N} = 2$ vectormultiplet which consists of a 4D $\mathcal{N} = 1$ vectormultiplet $V$ and
a chiral multiplet $\Sigma$.
Under the $Z_2$ parity, $V$ and $\Sigma$ are even and odd respectively.
The zero modes of $V$ are a massless gauge boson $A_\mu$ and a gaugino $\lambda$.
All the SM multiplets as well as particles of the messenger sector of gauge mediation
including vectorlike pairs of $\bf 5 + \bar{5}$, $\Psi$ and $\bar{\Psi}$,
are localized on the visible brane at $y = 0$.

\paragraph{Scherk-Schwarz SUSY breaking}
Let us now consider the Scherk-Schwarz mechanism of SUSY breaking.
The Scherk-Schwarz breaking is known to be equivalent to the radion $F$-term breaking
\cite{Marti:2001iw,Kaplan:2001cg} (for a recent discussion, see \cite{Craig:2014fka}).
The simplest model to get an $F$-term of the radion, $T = \pi R + \theta^2 F_T$,
is the no-scale model
\cite{Lahanas:1986uc}.\footnote{
For an example of the radion stabilization, see \cite{Luty:2002hj}.}
The effective 4D Lagrangian is given by
\begin{equation}
\begin{split}
\mathcal{L}_4^{\rm eff} &\supset - 3 M_5^3 \int d^4 \theta \, \phi^\dagger \phi \, (T + T^\dagger) \\[1ex]
&\quad + \int d^2 \theta \, \phi^3 \mathcal{C} + \rm h.c. \, ,
\end{split}
\end{equation}
where $\phi = 1 + \theta^2 F_\phi$ is the chiral compensator
and $\mathcal{C}$ is a constant superpotential with mass dimension 3 localized on one or both of the branes.
From this Lagrangian, we obtain
\begin{equation}
\begin{split}
F_T = \frac{\mathcal{C}^\ast}{M_5^3} \, , \qquad F_\phi = 0 \, ,
\label{Fterm}
\end{split}
\end{equation}
and a vanishing radion potential at tree-level.
Then, with a nonzero $\mathcal{C}$, SUSY is spontaneously broken and the cosmological constant is zero.
The fermion component of $T$ is the would-be goldstino absorbed by the gravitino.
The gravitino mass is given by
\begin{equation}
\begin{split}
m_{3/2} = \frac{\mathcal{C}^*}{M_4^2} = \frac{F_T}{L} \, .
\end{split}
\end{equation}
We take $m_{3/2} = \mathcal{O}(100) \, \rm TeV$ so that gauge mediation works as we will see below.
A small constant $\mathcal{C}$ compared to the naive value $M_5^3$ is due to an approximate $R$-symmetry.

Now, we transmit the SUSY breaking effect to the messenger sector. For this purpose, we consider the hidden $SU(N)$ in the bulk. The hidden gaugino zero mode gets a soft SUSY breaking mass by the coupling of the gauge field strength superfield with the radion $T$:
\begin{eqnarray}
\mathcal{L}_4^{\rm eff} \supset
\frac{1}{2} \left(\frac{F_T}{L} \right) \lambda_{\alpha}^a \lambda^{\alpha\, a} + \rm h.c.
\end{eqnarray}
This hidden gaugino mass induces a SUSY breaking $B$-term of the messenger superfields, which is required for gauge mediation, as shown below.
The hidden gauge interaction is asymptotically free below the compactification scale and finally confines.
We assume that the confinement scale is lower than the messenger mass scale.\footnote{
Although the gaugino condensation contributes to the superpotential,
the effect is tiny.
In fact, the $F$-term of the compensator $F_\phi$ is still very small and can be ignored.
}

\paragraph{The messenger sector}
Let us then focus on the messenger sector and consider the following superpotential
\cite{Nomura:2000uw,Nomura:1997ur}, 
\begin{eqnarray}
W = \lambda_1 X Q \bar Q + \lambda_m X \Psi \bar \Psi + \frac{\kappa}{3} X^3, \label{eq:super_mess}
\end{eqnarray}
where $X$ is a gauge singlet, $Q$ and $\bar Q$ are fundamental and anti-fundamental under the hidden $SU(N)$
and $\lambda_1$, $\lambda_m$, $\kappa$ are real coupling constants.
The above superpotential is protected by a $Z_{4R} \times Z_3$ symmetry,\footnote{
The $Z_3$ symmetry has $Z_3$-$SU(N)$-$SU(N)$ anomaly, which makes a domain wall unstable.
%
} where the $R$-charge of $X$ equals 2 and the other fields have vanishing $R$-charges. The $Z_3$ charges of the fields shown in Eq.(\ref{eq:super_mess}) are equal to 1, while the charges of the MSSM fields are 0.
From radiative corrections, $X$ has a negative soft mass and $X$ gets a vacuum expectation value (VEV) and an $F$-term, which becomes a source of gauge mediation. The radiative corrections can be estimated by the renormalization group running from the compactification scale to the messenger mass scale.  The relevant renormalization group equations 
are shown in Appendix A.
In Fig.~1, we show the renormalization group running of the gauge coupling and gaugino mass of the hidden $SU(N)$.

\begin{figure}[t]
\begin{center}
\includegraphics[scale=1.0]{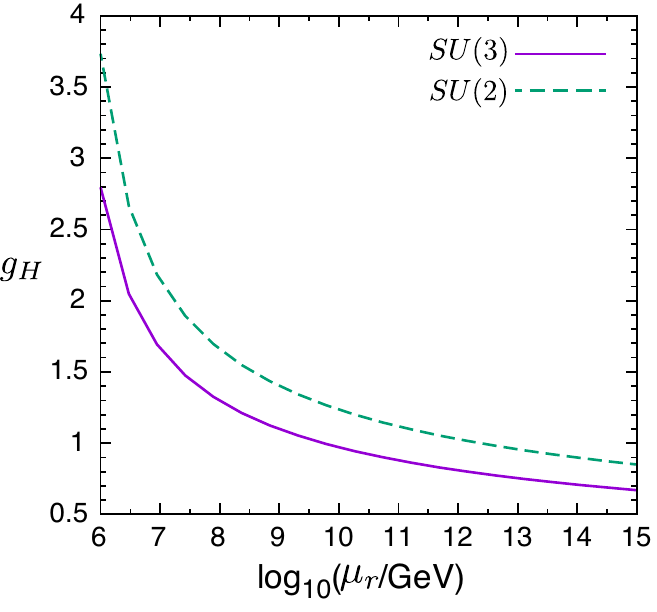}
\mbox{\raisebox{0.8mm}{\includegraphics[scale=1.0]{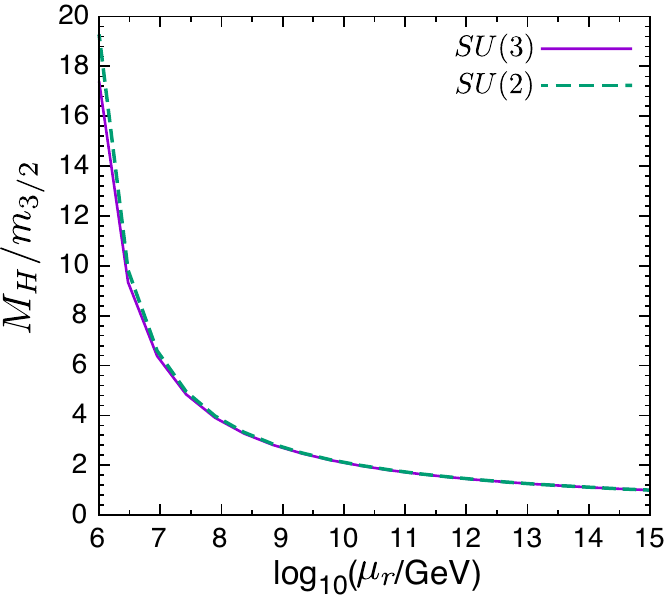}}}
\caption{The running of the hidden gauge coupling $g_H$ (left) and the gaugino mass $M_H$ normalized by $m_{3/2}$ (right). 
}
\label{fig:smallmu1}
\end{center}
\end{figure}

With positive soft masses squared for $Q$ and $\bar Q$ of $\sim 10^2\, m_{3/2}^2$, 
they are stabilized at the origin, and 
we have the mass term of the messenger scalars,
\begin{eqnarray}
V \supset \left(\kappa \frac{\left<X^\dag\right>^2}{\left<X\right>} +  A_m \right) \lambda_m \left<X\right>  \Psi \bar \Psi + {\rm h.c.},
\end{eqnarray}
where $\left<X\right>$ is determined by the scalar potential:
\begin{eqnarray}
V = -|m_X^2| |X|^2 + \kappa^2 |X|^4 + \frac{A_\kappa  \kappa}{3} X^3 + \rm h.c. ,
\end{eqnarray}
with $\sqrt{|m_X^2|} \sim A_{\kappa} \sim m_{3/2}$.
Here, $A_m$ and $A_\kappa$ are the $A$-terms of $X \Psi \bar{\Psi}$ and $X^3$ in the superpotential \eqref{eq:super_mess}.
The predicted spectrum is the same as that of the minimal gauge mediation, with identification of the parameters,
\begin{eqnarray}
M_{\rm mess} = \lambda_m \left<X\right>, \qquad \Lambda_m = \left(\kappa \frac{\left<X^\dag\right>^2}{\left<X\right>} +  A_m \right),
\end{eqnarray}
where $M_{\rm mess}$ is the messenger scale and $\Lambda_m \sim m_{3/2}$.
Numerically, we obtain $\Lambda_m \approx - 2.0\, m_{3/2}$ for a reasonable choice of the parameters, 
\begin{eqnarray}
(g_H=0.67, \, \lambda_1=0.7, \, \kappa=0.2, \, \lambda_m = 0.3) \ {\rm at} \  M_{\rm UV} = 10^{15}\,{\rm GeV},
\end{eqnarray}
where $M_{\rm UV}=L^{-1}$ and $N=3$ is taken.
Then, the SUSY breaking soft mass parameters for the MSSM particles are given by
\begin{eqnarray}
M_i = \frac{g^2_i}{16\pi^2} N_5 \Lambda_m, \qquad m_{\rm scalar}^2 = \frac{2N_5}{(16\pi^2)^2} \left( C_2^i g_i^4\right) |\Lambda_m|^2,
\end{eqnarray}
where $N_5$ is the number of vectorlike pairs of the messengers and $C_2^i$ ($i = 3,2,1$) is a quadratic Casimir for the SM gauge group.
Here, we stress again that anomaly mediation effects are suppressed because the $F$-term of the compensator $F_\phi$ is tiny
as in \eqref{Fterm}.
To obtain the Higgsino mass, we can write down the usual $\mu$-term in the superpotential on the visible brane.
By virtue of $F_\phi \approx 0$,
a too large $B_\mu$-term is not generated and
the correct $\mu$-$B_\mu$ relation is obtained in the SSGM framework.

\vspace{10pt}
Let us comment on a possible modification to the MSSM mass spectrum from gravitational interactions. The 5D gravitational multiplet in the bulk can mediate SUSY breaking of the radion $T$
into the ordinary sector.
The relevant one-loop effective K{\" a}hler potential is 
\cite{Rattazzi:2003rj}
\begin{equation}
\begin{split}
\Delta \mathcal{L}_4^{\rm eff} \sim \int d^4 \theta  \, \frac{1}{16 \pi^2} \frac{1}{(T + T^\dagger)^3 M_5^3} \, q^\dagger q  \, ,
\end{split}
\end{equation}
where $q$ denotes the SM matter fields collectively.
This term gives a universal contribution to the soft masses of the sfermions, and it does not cause dangerous FCNC processes. 
The contribution is proportional to the radion $F$-term,
\begin{equation}
\begin{split}
\Delta m_{\rm scalar}^2 &\sim 
\frac{1}{16 \pi^2} \left( \frac{1}{M_4 L} \right)^2 m_{3/2}^2 \, . \label{eq:rad_grav}
\end{split}
\end{equation}
This is sub-leading when $M_4 L > \mathcal{O}(100)$ is satisfied.
Therefore, gauge mediation gives the leading contribution to the soft masses of ordinary superparticles in the SSGM scenario.

\paragraph{Phenomenology}

As discussed above, the MSSM mass spectrum of this scenario is the same as the usual gauge mediation spectrum except the gravitino mass. Since the trilinear $A$ term of the top Yukawa is small in low-scale gauge mediation, relatively large squark masses ($\gtrsim 5$ TeV) are required to push the Higgs mass up in the minimal messenger sector case.

Sample mass spectra of such a simple messenger sector case are shown as the points {\bf I} and {\bf II} in Table~\ref{tab:sample}. 
For these points, we use {\tt SOFTSUSY 3.6.2}~\cite{softsusy} to calculate the MSSM mass spectra and {\tt FeynHiggs 2.11.2}~\cite{feynhiggs} for the Higgs mass calculation. As the minimal setup, we impose the condition that $B_\mu=0$ at the messenger scale, that is, the $\mu$-term is generated in a supersymmetric way. In this case, $\tan\beta$ is not a free parameter but a prediction. 
At both points, $\tan\beta$ is predicted to be large as $\sim 60$ and the stau is the LSP, which can be checked 
by charged track searches 
at future LHC experiments~\cite{hamaguchi, CMS:2013xfa}. 
At least, in such stau LSP cases, the stau needs to decay with a life-time shorter than $\mathcal{O}(1)$ second to avoid cosmological constraints. This can be done if $L L \bar E$ operators with small coefficients exist in the superpotential. An example of a realization is shown in Appendix B.
Note that relaxing the condition for the $B_\mu$-term allows the neutralino LSP. Furthermore, a non-minimal messenger sector can provide a variety of MSSM mass spectra. The neutralino LSP can be a candidate of the WIMP-like dark matter.  

The colored SUSY particles can be as light as 2-3 TeV, if one considers extensions of the MSSM.
One of the possible extensions is adding vector-like matter superfields, which enhances the Higgs boson mass via radiative corrections like the top/stop loops~\cite{moroi_okada, vector, vector2}. The MSSM mass spectra with the vector-like matter fields 
and the implications to LHC SUSY searches in a context of gauge mediation have been investigated extensively in Refs.~\cite{vgmsb, nakayama_yokozaki}, and it has been shown that there is a consistent solution with $B_\mu=0$ at the messenger scale~\cite{nakayama_yokozaki}.
At the point {\bf III} in Table~\ref{tab:sample}, the vector-like matter superfields which transform {\bf 10} and ${\bf \overline{10}}$ under the $SU(5)$ gauge group of the grand unified theory (GUT) are introduced as~\footnote{
Strictly speaking, the masses of the vector-like fields are split due to the renormalization group running from the GUT scale down to the weak scale. However, we ignore this effect for simplicity.
} 
\begin{eqnarray}
W \supset y' Q' H_u \bar U' + M_{10} (Q' \bar Q' + U' \bar U'  + E' \bar E'), 
\end{eqnarray}
where ${\bf 10}=(Q', \bar U', \bar E')$ and ${\bf \overline{10}} = (\bar Q', U', E')$.
In the calculation, we use {\tt Suspect} package~\cite{suspect} with appropriate modifications, and radiative corrections to the Higgs boson mass from the MSSM particles are evaluated using {\tt FeynHiggs 2.11.2}. The parameters, $y'$ and $M_{10}$, are taken as $y'=1.0$ and $M_{10}=1200$ GeV at the weak scale. 
Although we use the condition that $B_\mu=0$ at the messenger scale, the predicted value of $\tan\beta$ is not as large as those in the points {\bf I} and {\bf II} due to stronger interactions of the SM gauge group, 
and the LSP is the lightest neutralino. The SUSY mass scale is much smaller than the former two cases, which can be tested at the LHC.

%

The other usual extensions to obtain the $125$ GeV Higgs mass with a light sparticle spectrum in gauge mediation can also work. An investigation of the relation between such extensions and the WIMP dark matter possibility would be interesting, but it is beyond the scope of this paper.

\begin{table*}[]
\begin{center}
\begin{tabular}{|c||c|c|c|c|}
\hline
Parameters & Point {\bf I} & Point {\bf II} & Point {\bf III} \\
\hline
$\Lambda_{m}$ (TeV) & 400  & 600  & 220\\
$M_{\rm mess}$ (TeV)& 2000 & 2000  & 500 \\
$N_5$ & 2  & 1   & 1\\
\hline
$\tan\beta$ & 61.8 & 62.6 & 29.7 \\
$\mu$ (GeV) & 1907 & 1936  & 2220\\
\hline
%
Particles & Mass (GeV) & Mass (GeV) & Mass (GeV) \\
\hline
$\tilde{g}$ & 5260 & 4150  &2030 \\
$\tilde{q}$ & 5850 & 5840  & 3090-3190\\
$\tilde{t}_{2,1}$ & 5420, 5080 & 5010, 5380  & 3060, 2830\\
$\tilde{\chi}_{2,1}^\pm$ & 2110, 1900 & 1950, 1600 & 2230, 638 \\
$\tilde{\chi}_4^0$ & 2110 & 1950  & 2220 \\
$\tilde{\chi}_3^0$ & 1920 & 1940  & 2220\\
$\tilde{\chi}_2^0$ & 1900 & 1600  & 638\\
$\tilde{\chi}_1^0$ & 1110 & 842  &320 \\
$\tilde{e}_{L, R}(\tilde{\mu}_{L, R})$ & 2030, 1040 & 883, 409  & 883, 409\\
$\tilde{\tau}_{2,1}$ & 1970, 770 & 2025, 803 & 889, 360\\
$H^\pm$ & 686 & 643 &2390 \\
$h_{\rm SM\mathchar`-like}$ & 124.7 &  124.8 & 126.3\\
\hline
\end{tabular}
\caption{\small Mass spectra in sample points under the condition that $B_{\mu} = 0$ at the messenger scale.
For the point {\bf III}, the supersymmetric mass for the vector-like matter fields is taken to be $1200$ GeV.
}
\label{tab:sample}
\end{center}
\end{table*}

\section{Brane localized gravitational SUSY breaking}\label{sec:hiddenSUSY}

For gauge mediation with a very heavy gravitino, the Scherk-Schwarz mechanism is not necessarily required. 
Here we also propose another gauge mediation scenario with a heavy gravitino considering gravitational SUSY breaking~\cite{kiy,cp-safe} on the hidden brane at $y=\pi R$.

By assuming a shift symmetry of a SUSY breaking field $Z$ (which is slightly broken in the K{\" a}hler potential)~\cite{cp-safe}, the $Z$ dependence in the superpotential vanishes and the compensator $F$-term becomes zero as in the Scherk-Schwarz case.  
We consider the following 4D effective Lagrangian:
\begin{eqnarray}
\mathcal{L}_{4}^{\rm eff} = \int d^4 \theta \phi^\dag \phi \left[-3 M_4^2 + M_5^2 \tilde f(x) \right] + \int d^2 \theta \phi^3 \mathcal{C} 
+ \int d^2 \theta \phi^{\dag 3} \mathcal{C}^* ,
\end{eqnarray}
where $x= (Z+ Z^\dag)/M_5$, and the above Lagrangian is invariant under the shift, $Z \to Z + i \mathcal{R}$ ($\mathcal{R}$ is a real constant). Here, the first term comes from the radion contribution. The scalar potential is given by
\begin{equation}
\begin{split}
-V &= M_5^2 |F_\phi|^2 f + M_5^2 f' F_\phi^\dag F_x + M_5^2 f'  F_\phi F_x^\dag + M_5^2 f'' |F_x|^2 \\
&\quad+ 3 \mathcal{C} F_\phi  + 3 \mathcal{C}^* F_\phi^\dag,
\end{split}
\end{equation}
where $F_x = F_Z/M_5$ and $f(x) = -3 M_4^2/M_5^2 + \tilde f(x)$, and $f'$ denotes a derivative of $f$ in terms of $x$. Using the equations of motion of the $F$-terms, the potential reduces to a simple form as
$
V = - 3 \mathcal{C} F_\phi
$.
Then, the vanishing cosmological constant is obtained for $\left<F_\phi\right>=0$ or $\mathcal{C}=0
$. For $\mathcal{C} \neq 0$, SUSY is broken as follows. 
The $F$-terms are given by
\begin{equation}
\begin{split}
F_\phi =  \frac{3 \mathcal{C}^* f''}{M_5^2 (f'^2- f f'')} , \qquad
F_x =  -\frac{3 \mathcal{C}^*  f'}{M_5^2 (f'^2-  f f'')}.
\end{split}
\end{equation}
The apparent solution to $\left<F_\phi\right>=0$ is $\left<f''\right>=0$, then the SUSY is broken with $\left<F_Z\right>\neq0$. The minimum is found for 
$
\left<f^{(3)}(x)\right> = 0
$. For the canonically normalized $Z$, $|\left<F_Z\right>| = \sqrt{3} m_{3/2} M_4$.

As in the Scherk-Schwarz case, a hidden gauge multiplet propagates in the 5D space.
The hidden gaugino mass arises from the hidden brane localized coupling between $Z$ and the field strength superfield:
\begin{equation}
\begin{split}
\mathcal{L}_5 \supset  \int d^2 \theta \, \delta (y-\pi R) \frac{c_Z}{k} \frac{Z}{M_5^{2}}  \mathcal{W}_{\alpha}^a \mathcal{W}^{\alpha\, a} + {\rm h.c.}  
\, \to \, \mathcal{L}_4^{\rm eff} \supset  \frac{c_Z}{k}\frac{F_Z}{M_5^{2} L} \lambda_{\alpha}^a \lambda^{\alpha\,a} + {\rm h.c.},
\end{split}
\end{equation}
where $c_Z$ is a constant and $k=(\partial^2 K/\partial Z \partial Z^\dag)^{1/2}$ $[K=-3 M_4^2 \ln(-M_5^2 f/(3M_4^2))]$. Here, 
$\lambda_{\alpha}^a$ denotes the canonically normalized gaugino zero mode.
Then, the hidden gaugino gets a mass of $\sim m_{3/2}$ for $c_Z/k \sim 2$.
The messenger sector is the same as that in the Scherk-Schwarz case and
gauge mediation with a very heavy gravitino is realized.
Note that there is a radiative correction to the MSSM scalar masses from the exchange of the gravity multiplet in the bulk.
The effect turns out to be flavor universal and its size is very similar to Eq.(\ref{eq:rad_grav})~\cite{Rattazzi:2003rj}.

Finally, let us comment on the mass of the imaginary part of the SUSY breaking field $Z$. The shift-symmetry breaking in the K{\" a}hler potential may arise from radiative corrections as
\begin{eqnarray}
\Delta K = - \epsilon \frac{|Z|^4}{M_5^2},
\end{eqnarray}
where $\epsilon$ is a small constant.
This leads to $\Delta V \sim \epsilon\, |\left<F_Z\right>/M_5|^2  |Z|^2$, and the imaginary part gets a mass of $\sim \epsilon^{1/2} m_{3/2} (M_4/M_5)$ $\sim m_{3/2}$ for $\epsilon \sim 10^{-2}$.
Since the mass of the real part can be much larger than the gravitino mass due to the lower cutoff scale $M_5$, the cosmological moduli problem is also relaxed in this setup.

\section{Conclusion and discussions}\label{sec:conclusion}

In this paper, we propose a new framework of gauge mediation where the gravitino mass is $\mathcal{O}(100) \, \rm TeV$. We consider a flat extra dimension, a $S^1 / Z_2$ orbifold, and assume that all the SM multiplets as well as the messenger fields are localized on a brane at the fixed point. We consider two possibilities: 1) Scherk-Schwarz SUSY breaking and 2) gravitational SUSY breaking localized on the hidden brane as the SUSY breaking source, which is mediated to the messenger sector by radiative corrections from hidden gaugino loops.
In both scenarios, the compensator $F$-term can be substantially small and the anomaly mediation contribution can be ignored. The MSSM mass spectrum is, therefore, the same as that of usual gauge mediation except the gravitino mass in this scenario. Due to the vanishing $F$-term of the compensator field at the tree-level, the Higgs $B_\mu/\mu$ term is much smaller than the gravitino mass. The EWSB conditions are satisfied with,  for instance, vanishing $B_\mu$ 
at the messenger scale, taking into account radiative corrections from bino, wino and gluino loops.

The heavy gravitino, $m_{3/2} = \mathcal{O}(100) \, \rm TeV$, is cosmologically safe due to a short life-time, and it gives a possibility of the neutralino dark matter in gauge mediation. 
As in usual gauge mediation, the observed Higgs boson mass is obtained for the stop mass larger than about 5 TeV in the minimal model, but there would be a chance to detect the stau LSP as charged tracks at the LHC. Furthermore, usual extensions to enhance the Higgs mass in gauge mediation can also work. For example, as we have shown, the masses of the colored SUSY particles can be less than 2-3 TeV in the extension of the MSSM with vector-like matter fields. It is also the accessible range of future LHC experiments.

Finally, we comment on the cosmological moduli problem. 
In the gravitational SUSY breaking case, we have shown that the moduli can get masses around or heavier than the gravitino mass. In this case, the cosmological moduli problem is somewhat relaxed due to a relatively short life-time, but the overproduction of the gravitino is still problematic in general~\cite{moduli_induced}. Since subsequent decays of gravitinos may produce too many LSPs, the $R$-parity violation (or the equivalent interaction) may be required. Alternatively,
if the modulus coupling to the inflaton is strong, the overproduction of the LSP can be avoided~\cite{adiabatic1, adiabatic2}. 
In Scherk-Schwarz SUSY breaking, we assume a desirable stabilization mechanism of the radius of the extra dimension. However, in general, the moduli masses highly depend on the stabilization mechanism, which requires further investigation.

\paragraph{Note added}
While completing this manuscript, Ref.~\cite{note_added} was posted on arXiv, 
which has a very similar motivation but a different realization of gauge mediation.

\section{Acknowledgments}\label{sec:ackno}

This work is supported by the German Research Foundation through TRR33 ``The Dark Universe" (MA). YN is supported by a JSPS Fellowship for Research Abroad. The research leading to these results has received funding from the European Research Council under the European Unions Seventh Framework Programme (FP/2007-2013) / ERC Grant Agreement n. 279972 ``NPFlavour'' (NY).

\appendix
\section{Renormalization group equations}
The one-loop renormalization group equations of the parameters in the messenger sector are shown. 
Here, a pair of messenger superfields, which are ${\bf 5}$ and $\bar {\bf 5}$ in the $SU(5)$, is introduced. 
The contributions from the SM gauge interactions are neglected for simplicity.

{\allowdisplaybreaks  \begin{align} 
\beta_{g_H} & =   \frac{1}{16\pi^2}\left[
-(3N-1) g_H^{3} \right] \, , \\ 
\beta_{M_H} & =   \frac{1}{16\pi^2}\left[
-(6N-2) g_H^{2} M_H \right] \, , \\ 
\beta_{\lambda_m} & =  \frac{\lambda_m}{16\pi^2}\left[
 7 |\lambda_m|^2  +  2  |\kappa|^2  + N |\lambda_1|^2 
  \right] \, , \\ 
%
\beta_{\lambda_1} & =  \frac{\lambda_1}{16\pi^2}\left[ (N+2) |\lambda_1|^2 + 
2  |\kappa|^2    + 5  |\lambda_m|^2  -\frac{2(N^2-1)}{N} g_H^{2} \right]  \, , \\ 
\beta_{\kappa} & =  \frac{\kappa}{16\pi^2}\left[
6 |\kappa|^2  + 3 N |\lambda_1|^2  + 15 |\lambda_m|^2 \right] \, , \\ 
\beta_{A_{\lambda_m}} & =   \frac{1}{16\pi^2}\left[
14 |\lambda_m|^2 A_{\lambda_m} 
+ 4 |\kappa|^2 A_{\kappa} 
+ 2N |\lambda_1|^2 A_{\lambda_1}   \right] \, , \\ 
\beta_{A_{\lambda_1}} & =   \frac{1}{16\pi^2}\left[
(2N+4) |\lambda_1|^2 A_{\lambda_1} 
+ 4 |\kappa|^2 A_{\kappa}
+ 10 |\lambda_m|^2 A_{\lambda_m} 
+ \frac{4(N^2-1)}{N} g_H^{2} M_H \right] \, , \\ 
\beta_{A_{\kappa}} & =   \frac{1}{16\pi^2}\left[  12 |\kappa|^2 A_{\kappa} 
+ 6N |\lambda_1|^2 A_{\lambda_1}
+ 30 |\lambda_m|^2 A_{\lambda_m}\right] \, , \\ 
\beta_{m_{X}^2} & =   \frac{1}{16\pi^2}\left[
12 m_{X}^2 |\kappa|^2 +2 N \left(m_{Q}^2 + m_{\bar{Q}}^2 + m_{X}^2\right)|\lambda_1|^2 \right. \nonumber \\
& +10\left( m_{\Psi}^2 + m_{\bar{\Psi}}^2  + m_{X}^2 \right)|\lambda_m|^2 
\nonumber \\ 
 & \left. +4 |\kappa A_{\kappa}|^2 +2 N  |\lambda_1 A_{\lambda_1}|^2 
 +10 |\lambda_m A_{\lambda_m}|^2 \right] \, , \\ 
\beta_{m_{\Psi}^2} & =   \frac{1}{16\pi^2}\left[
 2 \left( m_{\Psi}^2 + m_{\bar{\Psi}}^2 + m_{X}^2 \right) |\lambda_m|^2  + 2 |\lambda_m A_{\lambda_m}|^2 \right] \, , \\ 
\beta_{m_{\bar{\Psi}}^2} & =  \beta_{m_{\Psi}^2}  \, , \\
\beta_{m_{Q}^2} & =  \frac{1}{16\pi^2}\left[
2 \left(m_{Q}^2 + m_{\bar{Q}}^2 + m_{X}^2\right)|\lambda_1|^2  + 2 |\lambda_1 A_{\lambda_1}|^2 -\frac{4(N^2-1)}{N} g_H^{2} |M_H|^2 \right] \, , \\ 
\beta_{m_{\bar{Q}}^2} & =   \beta_{m_{Q}^2} \, .
\end{align}} 
For the parameters presented here, see the main text.

\section{Consistent charge assignments for the (quasi) stable LSP}

\begin{table*}[!t]
\begin{center}
\begin{tabular}{|c|ccccc|}
\hline
 &  $H_u$ & $H_d$ & ${\bf 10}$ & ${\bf \bar{5}}$ & $\left<PQ\right>$   \\
 \hline
  $Z_{4R}$  &  $2$ & $0$ & $2$ & $0$ & $0$   \\
 \hline
  $U(1)_{\rm PQ}$  &  $2$ & $0$ & $-1$ & $1$ & $-1$   \\
\hline
\end{tabular}
\end{center}
\caption{An example of the charge assignment.
}
\label{tab:PQ}
\end{table*}

Here, we present  a charge assignment realizing the (quasi)stable LSP.
With the charge assignment shown in Table~\ref{tab:PQ}, the superpotential is 
\begin{eqnarray}
W &=& \xi_0 \frac{\left<PQ\right>^2}{M_5} H_u H_d +  {\bf 10\, 10} \,H_u + {\bf 10 \, \bar 5}\, H_d \nonumber \\
&+& \xi_1 \frac{\left<PQ\right>^3}{M_5^2} H_u {\bf \bar 5} + \xi_2 \frac{\left<PQ\right>}{M_5} {\bf 10\, \bar 5\, \bar 5}
,
\end{eqnarray}
where $\left<PQ\right> \sim 10^{10}$ GeV is the PQ breaking scale and $M_5 \sim 10^{17}$ GeV.
The dimension 5 proton decay operator, ${\bf 10\, 10\,  10\,  \bar 5}$, is highly suppressed by the $Z_{4R}$ and PQ-symmetry.

One can also consider a charge assignment, which reproduces ordinary $R$-parity conserving interactions, with
$R(H_u)=0$, $R(H_d)=2$ and $R({\bf 10})=3$, $R({\bf \bar 5})=1$. The PQ-breaking field, $\left<PQ\right>$, has a $R$-charge of $2$ in this case.

\end{document}